\documentclass[times,12pt]{article}

\usepackage{moreverb}
\usepackage{filecontents}
\usepackage{verbatim}
\usepackage[colorlinks,bookmarksopen,bookmarksnumbered,citecolor=red,urlcolor=red]{hyperref}
\usepackage{graphicx}
\usepackage{fancyvrb}

\makeatletter
\def\PY@reset{\let\PY@it=\relax \let\PY@bf=\relax%
    \let\PY@ul=\relax \let\PY@tc=\relax%
    \let\PY@bc=\relax \let\PY@ff=\relax}
\def\PY@tok#1{\csname PY@tok@#1\endcsname}
\def\PY@toks#1+{\ifx\relax#1\empty\else%
    \PY@tok{#1}\expandafter\PY@toks\fi}
\def\PY@do#1{\PY@bc{\PY@tc{\PY@ul{%
    \PY@it{\PY@bf{\PY@ff{#1}}}}}}}
\def\PY#1#2{\PY@reset\PY@toks#1+\relax+\PY@do{#2}}

\expandafter\def\csname PY@tok@gd\endcsname{\def\PY@tc##1{\textcolor[rgb]{0.63,0.00,0.00}{##1}}}
\expandafter\def\csname PY@tok@gu\endcsname{\let\PY@bf=\textbf\def\PY@tc##1{\textcolor[rgb]{0.50,0.00,0.50}{##1}}}
\expandafter\def\csname PY@tok@gt\endcsname{\def\PY@tc##1{\textcolor[rgb]{0.00,0.27,0.87}{##1}}}
\expandafter\def\csname PY@tok@gs\endcsname{\let\PY@bf=\textbf}
\expandafter\def\csname PY@tok@gr\endcsname{\def\PY@tc##1{\textcolor[rgb]{1.00,0.00,0.00}{##1}}}
\expandafter\def\csname PY@tok@cm\endcsname{\let\PY@it=\textit\def\PY@tc##1{\textcolor[rgb]{0.25,0.50,0.50}{##1}}}
\expandafter\def\csname PY@tok@vg\endcsname{\def\PY@tc##1{\textcolor[rgb]{0.10,0.09,0.49}{##1}}}
\expandafter\def\csname PY@tok@m\endcsname{\def\PY@tc##1{\textcolor[rgb]{0.40,0.40,0.40}{##1}}}
\expandafter\def\csname PY@tok@mh\endcsname{\def\PY@tc##1{\textcolor[rgb]{0.40,0.40,0.40}{##1}}}
\expandafter\def\csname PY@tok@go\endcsname{\def\PY@tc##1{\textcolor[rgb]{0.53,0.53,0.53}{##1}}}
\expandafter\def\csname PY@tok@ge\endcsname{\let\PY@it=\textit}
\expandafter\def\csname PY@tok@vc\endcsname{\def\PY@tc##1{\textcolor[rgb]{0.10,0.09,0.49}{##1}}}
\expandafter\def\csname PY@tok@il\endcsname{\def\PY@tc##1{\textcolor[rgb]{0.40,0.40,0.40}{##1}}}
\expandafter\def\csname PY@tok@cs\endcsname{\let\PY@it=\textit\def\PY@tc##1{\textcolor[rgb]{0.25,0.50,0.50}{##1}}}
\expandafter\def\csname PY@tok@cp\endcsname{\def\PY@tc##1{\textcolor[rgb]{0.74,0.48,0.00}{##1}}}
\expandafter\def\csname PY@tok@gi\endcsname{\def\PY@tc##1{\textcolor[rgb]{0.00,0.63,0.00}{##1}}}
\expandafter\def\csname PY@tok@gh\endcsname{\let\PY@bf=\textbf\def\PY@tc##1{\textcolor[rgb]{0.00,0.00,0.50}{##1}}}
\expandafter\def\csname PY@tok@ni\endcsname{\let\PY@bf=\textbf\def\PY@tc##1{\textcolor[rgb]{0.60,0.60,0.60}{##1}}}
\expandafter\def\csname PY@tok@nl\endcsname{\def\PY@tc##1{\textcolor[rgb]{0.63,0.63,0.00}{##1}}}
\expandafter\def\csname PY@tok@nn\endcsname{\let\PY@bf=\textbf\def\PY@tc##1{\textcolor[rgb]{0.00,0.00,1.00}{##1}}}
\expandafter\def\csname PY@tok@no\endcsname{\def\PY@tc##1{\textcolor[rgb]{0.53,0.00,0.00}{##1}}}
\expandafter\def\csname PY@tok@na\endcsname{\def\PY@tc##1{\textcolor[rgb]{0.49,0.56,0.16}{##1}}}
\expandafter\def\csname PY@tok@nb\endcsname{\def\PY@tc##1{\textcolor[rgb]{0.00,0.50,0.00}{##1}}}
\expandafter\def\csname PY@tok@nc\endcsname{\let\PY@bf=\textbf\def\PY@tc##1{\textcolor[rgb]{0.00,0.00,1.00}{##1}}}
\expandafter\def\csname PY@tok@nd\endcsname{\def\PY@tc##1{\textcolor[rgb]{0.67,0.13,1.00}{##1}}}
\expandafter\def\csname PY@tok@ne\endcsname{\let\PY@bf=\textbf\def\PY@tc##1{\textcolor[rgb]{0.82,0.25,0.23}{##1}}}
\expandafter\def\csname PY@tok@nf\endcsname{\def\PY@tc##1{\textcolor[rgb]{0.00,0.00,1.00}{##1}}}
\expandafter\def\csname PY@tok@si\endcsname{\let\PY@bf=\textbf\def\PY@tc##1{\textcolor[rgb]{0.73,0.40,0.53}{##1}}}
\expandafter\def\csname PY@tok@s2\endcsname{\def\PY@tc##1{\textcolor[rgb]{0.73,0.13,0.13}{##1}}}
\expandafter\def\csname PY@tok@vi\endcsname{\def\PY@tc##1{\textcolor[rgb]{0.10,0.09,0.49}{##1}}}
\expandafter\def\csname PY@tok@nt\endcsname{\let\PY@bf=\textbf\def\PY@tc##1{\textcolor[rgb]{0.00,0.50,0.00}{##1}}}
\expandafter\def\csname PY@tok@nv\endcsname{\def\PY@tc##1{\textcolor[rgb]{0.10,0.09,0.49}{##1}}}
\expandafter\def\csname PY@tok@s1\endcsname{\def\PY@tc##1{\textcolor[rgb]{0.73,0.13,0.13}{##1}}}
\expandafter\def\csname PY@tok@kd\endcsname{\let\PY@bf=\textbf\def\PY@tc##1{\textcolor[rgb]{0.00,0.50,0.00}{##1}}}
\expandafter\def\csname PY@tok@sh\endcsname{\def\PY@tc##1{\textcolor[rgb]{0.73,0.13,0.13}{##1}}}
\expandafter\def\csname PY@tok@sc\endcsname{\def\PY@tc##1{\textcolor[rgb]{0.73,0.13,0.13}{##1}}}
\expandafter\def\csname PY@tok@sx\endcsname{\def\PY@tc##1{\textcolor[rgb]{0.00,0.50,0.00}{##1}}}
\expandafter\def\csname PY@tok@bp\endcsname{\def\PY@tc##1{\textcolor[rgb]{0.00,0.50,0.00}{##1}}}
\expandafter\def\csname PY@tok@c1\endcsname{\let\PY@it=\textit\def\PY@tc##1{\textcolor[rgb]{0.25,0.50,0.50}{##1}}}
\expandafter\def\csname PY@tok@kc\endcsname{\let\PY@bf=\textbf\def\PY@tc##1{\textcolor[rgb]{0.00,0.50,0.00}{##1}}}
\expandafter\def\csname PY@tok@c\endcsname{\let\PY@it=\textit\def\PY@tc##1{\textcolor[rgb]{0.25,0.50,0.50}{##1}}}
\expandafter\def\csname PY@tok@mf\endcsname{\def\PY@tc##1{\textcolor[rgb]{0.40,0.40,0.40}{##1}}}
\expandafter\def\csname PY@tok@err\endcsname{\def\PY@bc##1{\setlength{\fboxsep}{0pt}\fcolorbox[rgb]{1.00,0.00,0.00}{1,1,1}{\strut ##1}}}
\expandafter\def\csname PY@tok@mb\endcsname{\def\PY@tc##1{\textcolor[rgb]{0.40,0.40,0.40}{##1}}}
\expandafter\def\csname PY@tok@ss\endcsname{\def\PY@tc##1{\textcolor[rgb]{0.10,0.09,0.49}{##1}}}
\expandafter\def\csname PY@tok@sr\endcsname{\def\PY@tc##1{\textcolor[rgb]{0.73,0.40,0.53}{##1}}}
\expandafter\def\csname PY@tok@mo\endcsname{\def\PY@tc##1{\textcolor[rgb]{0.40,0.40,0.40}{##1}}}
\expandafter\def\csname PY@tok@kn\endcsname{\let\PY@bf=\textbf\def\PY@tc##1{\textcolor[rgb]{0.00,0.50,0.00}{##1}}}
\expandafter\def\csname PY@tok@mi\endcsname{\def\PY@tc##1{\textcolor[rgb]{0.40,0.40,0.40}{##1}}}
\expandafter\def\csname PY@tok@gp\endcsname{\let\PY@bf=\textbf\def\PY@tc##1{\textcolor[rgb]{0.00,0.00,0.50}{##1}}}
\expandafter\def\csname PY@tok@o\endcsname{\def\PY@tc##1{\textcolor[rgb]{0.40,0.40,0.40}{##1}}}
\expandafter\def\csname PY@tok@kr\endcsname{\let\PY@bf=\textbf\def\PY@tc##1{\textcolor[rgb]{0.00,0.50,0.00}{##1}}}
\expandafter\def\csname PY@tok@s\endcsname{\def\PY@tc##1{\textcolor[rgb]{0.73,0.13,0.13}{##1}}}
\expandafter\def\csname PY@tok@kp\endcsname{\def\PY@tc##1{\textcolor[rgb]{0.00,0.50,0.00}{##1}}}
\expandafter\def\csname PY@tok@w\endcsname{\def\PY@tc##1{\textcolor[rgb]{0.73,0.73,0.73}{##1}}}
\expandafter\def\csname PY@tok@kt\endcsname{\def\PY@tc##1{\textcolor[rgb]{0.69,0.00,0.25}{##1}}}
\expandafter\def\csname PY@tok@ow\endcsname{\let\PY@bf=\textbf\def\PY@tc##1{\textcolor[rgb]{0.67,0.13,1.00}{##1}}}
\expandafter\def\csname PY@tok@sb\endcsname{\def\PY@tc##1{\textcolor[rgb]{0.73,0.13,0.13}{##1}}}
\expandafter\def\csname PY@tok@k\endcsname{\let\PY@bf=\textbf\def\PY@tc##1{\textcolor[rgb]{0.00,0.50,0.00}{##1}}}
\expandafter\def\csname PY@tok@se\endcsname{\let\PY@bf=\textbf\def\PY@tc##1{\textcolor[rgb]{0.73,0.40,0.13}{##1}}}
\expandafter\def\csname PY@tok@sd\endcsname{\let\PY@it=\textit\def\PY@tc##1{\textcolor[rgb]{0.73,0.13,0.13}{##1}}}

\def\PYZbs{\char`\\}
\def\PYZus{\char`\_}
\def\PYZob{\char`\{}
\def\PYZcb{\char`\}}

\def\PYZlt{\char`\<}
\def\PYZgt{\char`\>}
\def\PYZsh{\char`\#}

\def\PYZhy{\char`\-}
\def\PYZsq{\char`\'}

\makeatother

\newcommand\BibTeX{{\rmfamily B\kern-.05em \textsc{i\kern-.025em b}\kern-.08em
T\kern-.1667em\lower.7ex\hbox{E}\kern-.125emX}}

\begin{document}


\title{Proof-Driven Development}

\author{B. Goodspeed}



\begin{abstract}
A new workflow for software development (proof-driven development) is presented.  An extension of test-driven development, the new workflow utilizes the paradigm of dependently typed programming.  The differences in design, complexity and provability of software are discussed, based on the technique used to create the system.  Furthermore, the difference in what properties can be expressed in a proof-driven development workflow versus a traditional test-driven development workflow or using test-last development.
\end{abstract}


\maketitle

\section{Introduction}
Program correctness and software quality poses a serious concern in our lives.  We not only carry computing devices on our person, but in our person.  Many advances have been made recently to the way we construct software, including agile methodologies like test-driven development.  Our programming languages have been evolving as well, giving us new paradigms we are only just learning to leverage.  Several of these languages can be used for formal mathematical proof, and support extremely strong type checking at compile time.

The earlier a defect is detected, the easier it is to resolve.  Finding defects with a compiler means they are never introduced into a completed build.  We can never fix every defect \cite{Loscocco:InevitabilityFailure:01}.  Even mathematics cannot save us entirely from the complexity of the systems we build \cite{Raatikainen:Goedel:13}.   However, we can reduce the number of defects we allow \cite{Maximilien:TDDDefectRates:03}.   Frequently, the tools used to reduce such defect rates are mathematical proofs of correctness \cite{Leroy:FormalCompiler:09, Klean:FormalKernel:09, Jang:BrowserSecurity:12, Chlipala:DependentTypes:08, Appel:VerifiedSHA256:14, Almeida:CertifiedCrypto:13}, and improvements to the processes used to create software \cite{Beck:TDD:03, Beznosov:AgileSecAssurance:04, Feathers:WEWLC:04, Fowler:Refactoring:99, Martin:CleanCode:08, Prowell:CleanRoom:99, Schwaber:SCRUM:01}.

This work proposes a merging of these two families of ideas.

\subsection{Gaps}
\label{gaps}
When we discuss formally (mathematically) verified software, we find the workflow (which we refer to as the ``common workflow", below) with the following stages has been broadly employed by the papers cited here.
\begin{enumerate}
\setlength{\itemsep}{0pt}
\setlength{\parskip}{0pt}
\setlength{\parsep}{0pt}    
\item Define security goals.
\item Produce specification or policy.
\item Produce model.
\item Prove the model meets the goals.
\item Implement a system based on the model.
\end{enumerate}
\label{workflow}

In theory, the ``common workflow" should suffice.  But, there is a gap between theory and practice (or ``in theory, theory is practice; in practice it's not."\footnote{Variations of this quote have been attributed to DaVinci, Fermat, Pascal and Einstein.}).   In particular, a problematic gap exists where a model (about which proofs exist) is used only as a guide to the implementation.  A painful example of this was the recent Heartbleed bug \cite{Wheeler:PreventingHeartbleed:14} in an extension to the open source cryptography library OpenSSL.  While the cryptographic base framework was proven to be sound (at least as far as conjectures surrounding the difficulty of mathematical problems like integer factorization and the discrete log problem allow), the implementation was not sound.  To produce secure systems more reliably, we must reduce the number and size of such gaps.

In this paper, we provide the following central contribution:
\begin{itemize}
\setlength{\itemsep}{0pt}
\setlength{\parskip}{0pt}
\setlength{\parsep}{0pt}    
\item We introduce an alternative to the common workflow, which we call ``Proof driven development" (PDD).  Proof-driven development represents an extension to existing test-driven development workflows, combined with new technology made available by proof assistants and dependently typed languages.  This contribution is described in section \ref{chapter:PDD}.
\end{itemize}

\subsection{Language Selection}
As proof-related features are somewhat esoteric in the realm of programming languages, we can begin our search with tools/systems that self-identify as proof assistants\footnote{http://en.wikipedia.org/wiki/Proof\_assistant}.  We select our language based on two key characteristics:
\begin{itemize}
\item the language's power as a theorem prover, and;  
 
\item the ability to extract executable machine code.
\end{itemize}

To consider a system a theorem prover for our discussions, we need at least universal quantification over a predicate (e.g. $\forall x, P(x) = true$) rather than simple existential quantification (e.g. $\exists x, P(x)=true$).  Some of the implications of this broader ability are discussed in section \ref{chapter:PDD}.

Even in systems that support the extraction of runnable code, the extraction process is not always built-in.  For example, Coq \cite{Bertot:CoqHurry:10} has the ability to extract ML or Haskell code which can then be made executable by the toolchains provided by Haskell and ML.  However, the extracted code often loses many of the useful type-safety properties.  Some issues related to extraction are described by Swierstra in his experience report \cite{Swierstra:Extractions:12} on working with ML code extracted from Coq.  

As some tools do not support either compilation or extraction,  these tools will never be suitable for end-to-end software construction.

We have chosen the language Idris \cite{Brady:IdrisSystemsProgramming:11} for our demonstration of the new proof-driven development workflow.  However, other dependently typed languages such as Cayenne \cite{Augustsson:Cayenne:99}, Agda \cite{Norell:Agda:09}, or Coq \cite{Bertot:CoqArt:04} would suffice in terms of quantification power.  In the words of the designer of Idris, it is a ``systems programming language with dependent types" \cite{Brady:IdrisSystemsProgramming:11}, and thus suitable for practical programming.  It can be thought of as a combination of Agda and Haskell.

By design, Idris features: 
\begin{quote}
\textit{
``easy interoperability with C and high level language constructs to support domain specific language implementation. Idris emphasises general-purpose programming, rather than theorem proving, and as such includes higher level programming constructs such as type classes and `do-notation'. Idris also supports tactic based theorem proving, and has a lightweight Hugs/GHCI style interface." \footnote{Idris FAQ: http://docs.idris-lang.org/en/latest/faq/faq.html}
}
\end{quote}

\section{Proof Driven Development}

\label{chapter:PDD}

The software development lifecycle (SDLC), for both ``normal" software, and ``mission critical" software requiring the utmost security includes: specification, design, implementation and testing.  However, there are many schools of thought concerning the ``best" way to organize the progress between these states during the construction of software.  In waterfall development (described in detail in section \ref{Waterfall}), the states are not revisited, but progressed through linearly.  In so-called ``Agile" processes (see section \ref{Agile}), such as ``eXtreme Programming" (XP) \cite{Beck:XPExplained:99}, these states are visited repeatedly during the construction of software, in a spiral pattern.  These two schools of thought about the construction of software tend to apply at different times.  Waterfall is well-suited to problem domains where the issues are well-understood, and new unknowns are not likely to arise during construction.  Agile is better suited to exploration and research, as it does not make the assumption that all the facts are in before implementation begins, and permits requirements to change during the construction process.  

This workflow is pictured in figure \ref{fig:OriginalNumberedWorkflow}.  The numbers for each stage in the figure refer to the artifacts, showing clearly the flow from specifications to mental model to mathematical model to source code.  The source code is then compiled into the program, and finally the proofs are written.


Here, we make an argument for a mixed approach to secure software construction, as there are important lessons to be learned from both approaches.
\section{Waterfall}
\label{Waterfall}


Waterfall assumes that all design decisions can be made before the software/system is constructed.  This tends to apply in well-understood areas.  Waterfall grew out of very early attempts to formalize software construction processes \cite{Benington:Waterfall:83}.  These early approaches were based on electrical and hardware engineering techniques.  The authors admitted that it was not an ideal approach due to the differences between these engineering disciplines (being well understood) and the mathematically/algorithmically oriented software engineering discipline which was only just emerging.

Many Agile adherents \cite{Beznosov:AgileSecAssurance:04, Martin:CleanCode:08, Schwaber:SCRUM:01, Beck:XPExplained:99} argue that waterfall is costly and unrealistic.  However, pre-loading the design and specification phases is very much what tends to happen with secure software systems, especially considering the academic sources of many designs/specifications.  Many of the specifications/systems have been proven on paper and published without the original authors intending to implement the systems themselves.  This allows for a great deal of caution and review from many experts.  Indeed, the most common workflow (described in section \ref{gaps} and figure \ref{fig:OriginalNumberedWorkflow}) is based on this idea.  

However, the criticisms of waterfall in terms of emerging issues, and the speed with which the final system is available are valid, and it is clear that the waterfall approach by itself is not sufficient.

\section{Agile/TDD}
\label{Agile}

Agile itself is a blanket term for many software development processes including Clean Room \cite{Prowell:CleanRoom:99} -- which incorporates the use of formal methods such as model checking, process algebras, and testing, XP \cite{Beck:XPExplained:99}, and SCRUM \cite{Schwaber:SCRUM:01}.  The core values of Agile systems are embodied by their manifesto:
\begin{quote}
\textit{
We are uncovering better ways of developing
software by doing it and helping others do it.
Through this work we have come to value:
\begin{itemize}
\item Individuals and interactions over processes and tools
\item Working software over comprehensive documentation
\item Customer collaboration over contract negotiation
\item Responding to change over following a plan
\end{itemize}
That is, while there is value in the items on
the right, we value the items on the left more.
}
\footnote{http://agilemanifesto.org/}
\end{quote}

Clearly, there is a large variety of so-called Agile methods, and they all vary from one to the next.  However, one key discipline, notable in a great deal of these methods,  is called ``test-driven development" or TDD.  Test-driven development was described in great detail by Kent Beck in his introductory guide to TDD \cite{Beck:TDD:03}, and it is described in a great deal of related texts \cite{Fowler:Refactoring:99,  Martin:CleanCode:08, Maximilien:TDDDefectRates:03, Beznosov:AgileSecAssurance:04, Schwaber:SCRUM:01}.

Test-driven development refers to the reversal of the ``common" workflow for creating software.  Instead of writing a piece of software, and then writing a test to determine if the software is working correctly, the test is created first to ``drive" the creation of the software required to satisfy the assertions in the test.
This is important for several reasons, described in detail by Beck and Martin in their works on software craftsmanship\cite{Beck:XPExplained:99, Beck:TDD:03, Martin:CleanCode:08}.  First, budget and timeline pressure often forces things on the end of the spectrum to be dropped (that is: the tests never get written).  Second, the tests ``driving" (or ``pulling", in Toyota Production \cite{Ono:Toyota:88} or Lean \cite{Poppendieck:Lean:03} terminology) the functionality help to keep the system minimal (``lean"), with no unnecessary functions.  Third, the creation of tests up-front force the developers to think of the users of the code (be it other systems or end-users).  Finally, the tests form a suite of properties that can be automatically verified, thus helping to avoid regression errors  (re-introducing bugs that were previously fixed into new releases).

While TDD has gained substantial traction in recent years, with dozens of publications extolling its virtues, the idea is not new.  Anecdotally\footnote{http://c2.com/cgi/wiki?TenYearsOfTestDrivenDevelopment}, the Mercury Space Program at NASA used a variant of TDD with their punch-card programming system: punching the cards with the expected output, then programming the system until the output punchcards matched.  What has changed recently is the availability and quality of tools built to support this new workflow.  Many frameworks are based on early work by Beck \cite{Beck:TDD:03}, who created the first ``xUnit" unit testing framework, SUnit for Smalltalk.  SUnit (and Smalltalk itself) did not gain a huge marketshare (below 0.21\% according to TIOBE \footnote{http://www.tiobe.com/index.php/content/paperinfo/tpci/index.html}).  However, JUnit (a Java version SUnit) saw a great deal of success - with many books discussing it, including Beck's and Martin's well known volumes on practical software construction \cite{Beck:TDD:03, Martin:CleanCode:08}.  These frameworks made TDD approachable to industry, and promoted the ``red-green-refactor" workflow of creating a failing test case (red), making the test case pass by creating the simplest code possible (green), then refactoring the code to make it clean \cite{Fowler:Refactoring:99}.  The workflow has had a remarkable effect on defect rates, with a study at IBM showing a 50\% reduction over ad-hoc testing \cite{Maximilien:TDDDefectRates:03}.

Testing and test coverage (the proportion of code-paths covered by tests) are influenced by the source language, and the complexity \cite{McCabe:SoftwareComplexity:76} of the methods/classes under test.  The limiting factor is that fundamentally all tests are fixed-point assertions.  That is, for a function $f$, we can specify the output for any fixed input $y = f(x)$.  We can do this for arbitrarily many values of $x$, but it is still finite instances of existential quantification (that there is an $x$ for which this function behaves correctly).  Extending this idea with statistical sampling, we arrive at the work of Claessen and Hughes, QuickCheck \cite{Claessen:Quickcheck:11}.  Their work is based on the idea that any value from a range of inputs should produce valid output within another well-defined range (and not crash).  The system uses these ranges as inputs and statistically samples values within them (the number of which is configurable) and synthesizes new assertions/test cases.  This is analogous to how a human would extend the same tests, but without the tedium of creating them manually.  

Even with such statistical sampling, we cannot say the function is correct for any input.  In formal terms we cannot state ``$\forall x, f(x)$ is valid."  Section \ref{ExampleExistentialProblem} shows why this is still a source of defects.

\subsection{Example Existential Problems}

\label{ExampleExistentialProblem}
To illustrate why testing based on ``arbitrarily many" existential inputs is not sufficient to cover all error cases that a universally quantified assertion would cover, we present the following example.  In mathematical terms, we define a relation $f(x) = x/x$ where the inputs and outputs of $f$ are the Real numbers.  
In code, this could be implemented as shown in figure \ref{fig:CProgramWithBug}.
\begin{figure}[h]
\begin{Verbatim}[commandchars=\\\{\}]
\PY{k+kt}{double} \PY{n+nf}{f}\PY{p}{(}\PY{k+kt}{double} \PY{n}{x}\PY{p}{)} \PY{p}{\PYZob{}}
    \PY{k}{return} \PY{n}{x}\PY{o}{/}\PY{n}{x}\PY{p}{;}
\PY{p}{\PYZcb{}}
\end{Verbatim}
\caption{C program with bug.}\label{fig:CProgramWithBug}
\end{figure}

\begin{figure}[h]
\includegraphics[width=0.5\textwidth]{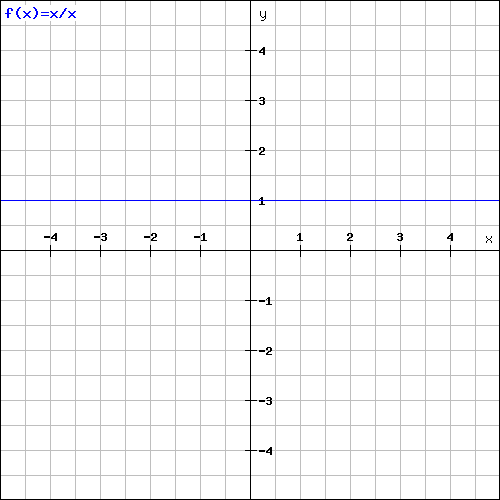} 
\caption{Graph with hole.}\label{fig:HoleInGraph}
\end{figure}

Plotting this function, we would see the graph in figure \ref{fig:HoleInGraph}.

For arbitrarily many values this method gives the correct and expected return value, $1$.  However, for the input $0$, this program causes a division by zero error.  So, not only does one value (just one, out of as many distinct values as the data type supports) give an unexpected result, this input actually crashes the program.  While this is a contrived example, the issue exposed here is very real.  

It is not only numerical issues that can cause unexpected bugs during translation from symbolic mathematics into computational datatypes and functions.  Other datatypes suffer from related problems.  Take for example the Java method in figure \ref{fig:JavaWithBug}. 
\begin{figure}[h]
\begin{Verbatim}[commandchars=\\\{\}]
\PY{n}{String} \PY{n+nf}{convertBoolean}\PY{o}{(}\PY{n}{Boolean} \PY{n}{b}\PY{o}{)} \PY{o}{\PYZob{}}
       \PY{k}{return} \PY{n}{b}\PY{o}{.}\PY{n+na}{toString}\PY{o}{(}\PY{o}{)}\PY{o}{;}
\PY{o}{\PYZcb{}}
\end{Verbatim}

\caption{Java program with bug.}\label{fig:JavaWithBug}
\end{figure}

It would seem that by testing the function with the values \textit{true} and \textit{false}, we would exhaustively test this function.  However, the Boolean data type in Java is actually tri-state.  It is possible to crash this program by passing \textit{null}, a Boolean (distinct from the primitive boolean data type) value that is neither true nor false.  To be truly confident in our reasoning about the behavior of functions we need to be able to assert behavior for all possible values of our data types.

\section{PDD}

\begin{figure}[h]
\includegraphics[width=\textwidth]{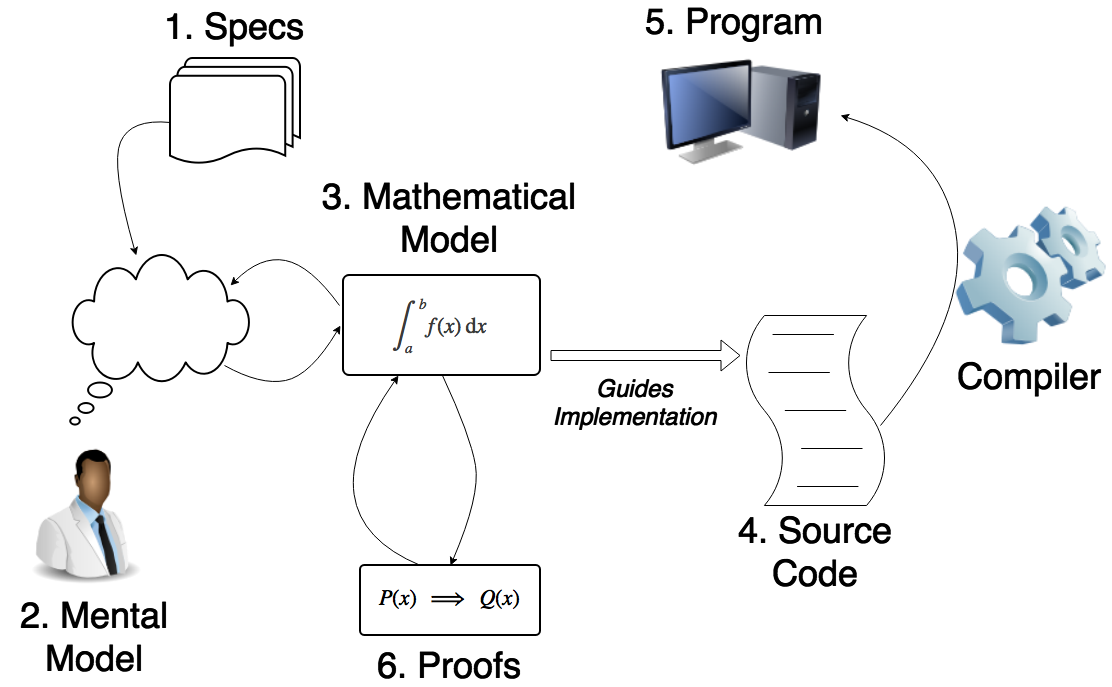} 
\caption{Common workflow (numbered).}\label{fig:OriginalNumberedWorkflow}
\end{figure}

In order to close gaps in the common workflow described in section \ref{gaps}, as shown in figure \ref{fig:OriginalNumberedWorkflow}, we extend the notion of test driven development and Clean Room software engineering with what we dub ``proof-driven development", PDD.  The term has been used before by Bird to describe a similar technique used to iteratively refine an algorithm to solve Sudoku puzzles \cite{Bird:ProofSudoku:06}.  However, the proofs described therein are not machine verified (essentially Bird is talking more about proof-backed refactoring).  Using proof-driven development techniques without automation leaves the system (and proofs) vulnerable to regression errors.  Regression errors are a common problem in areas that require change or evolution (as with nearly all software systems). 

\begin{figure}[h]
\includegraphics[width=\textwidth]{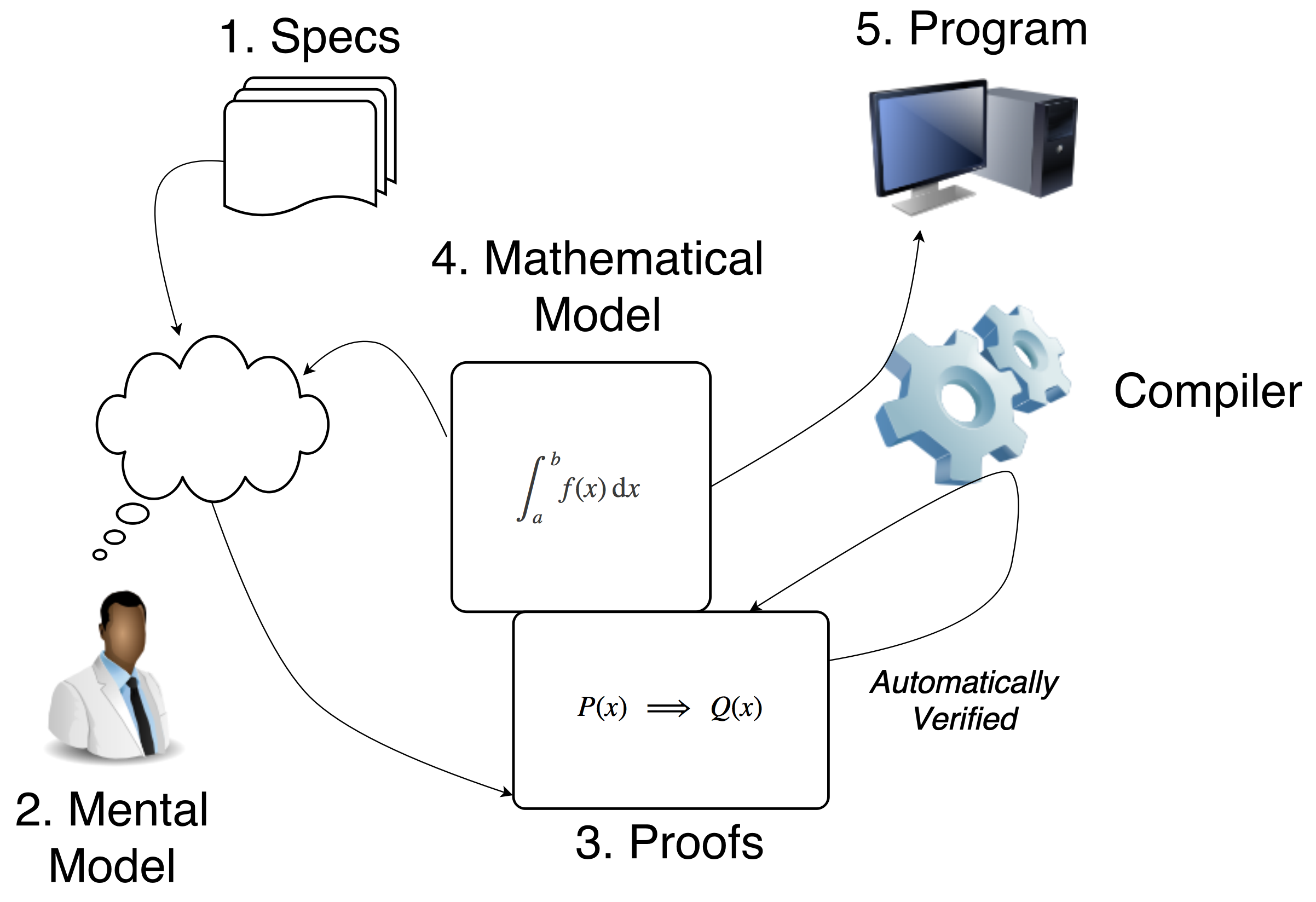} 
\caption{Improved workflow (numbered).}\label{fig:ImprovedNumberedWorkflow}
\end{figure}

Using the power of dependent types in a proof assistant language (for our purposes, Idris), we are able to perform specifications with universal quantification.  Furthermore, by selecting a language with support for extraction/compilation, we close an entire problematic gap in the common workflow diagram (figure \ref{fig:OriginalNumberedWorkflow}).  The improved workflow is shown in figure \ref{fig:ImprovedNumberedWorkflow}.  The first two steps, creating specifications and synthesizing a mental model, remain the same as in the common workflow.  However, the next step is to create proof statements of desired properties.  These proofs (and the output from the compiler/verifier) guide the creation of the mathematical model.  The mathematical model takes the place of the extracted/derived source code, and the compiler directly produces a program.  

As with TDD, success will depend on the quality of the framework and language upon which the system is based.  When we make a proof-based, universally quantified assertion, such as ``$\forall x, f(x)$ does not crash", we need our system to tell us if in fact we have changed $f$ such that it is not longer true.  Ideally during development, the system will guide us/the user to the ``correct" data types to complete the proof of correctness.  

Systems like Agda and Idris that support ``holes", where the compiler can tell what type of computation is needed to get from state A to state B, are close.  Similarly, systems with proof search such as Coq, Agda, and Idris, can make use of the compiler information about the hole in the proof.

To illustrate this, consider the line of reasoning about transitivity of homomorphisms in Idris in figure \ref{fig:Equational}.  This is a statement and partial proof of transitivity of semigroup homomorphisms (details and full sources online\footnote{http://github.com/bgoodspeed/idris-misc}).  It states that two homomorphisms $h$ and $h'$ can be evaluated in either order.  The proof steps are elided by the holes ``?prf1" and ``?prf2".  The system can be queried as to the exact type required for each hole (which in this case is a complex expression due to the dictionary tracking, necessary for distinguishing which semigroup operation is being referenced).

This signature can then be passed to the proof search engine.  For example, if we needed a transformation from Nat (the type denoting natural numbers in Idris) to Bool (the type denoting boolean values in Idris), we could search as shown in figure \ref{fig:ReplSearch}.  The results of which show us that the system knows two such predicates, isSucc (a decision property to determine if a given natural number is a successor form (greater than zero), and isZero, which matches if the natural number is zero.

\begin{figure}[h]
\begin{Verbatim}[commandchars=\\\{\}]
\PY{n+nf}{homTrans} \PY{o+ow}{:} \PY{o+ow}{(}adict \PY{o+ow}{:} \PY{k+kt}{Semigroup} a, bdict \PY{o+ow}{:} \PY{k+kt}{Semigroup} b, 
       cdict \PY{o+ow}{:} \PY{k+kt}{Semigroup} c\PY{o+ow}{)} \PY{o+ow}{=\PYZgt{}}
           \PY{k+kt}{Hom} a b adict bdict \PY{o+ow}{\PYZhy{}\PYZgt{}} \PY{k+kt}{Hom} b c bdict cdict 
           \PY{o+ow}{\PYZhy{}\PYZgt{}} \PY{k+kt}{Hom} a c adict cdict
homTrans \PY{o+ow}{@\PYZob{}}adict\PY{o+ow}{\PYZcb{}} \PY{o+ow}{@\PYZob{}}bdict\PY{o+ow}{\PYZcb{}} \PY{o+ow}{@\PYZob{}}cdict\PY{o+ow}{\PYZcb{}} \PY{o+ow}{(}\PY{k+kt}{MkHom} h preservesGroup\PY{o+ow}{)} 
        \PY{o+ow}{(}\PY{k+kt}{MkHom} h\PYZsq{} preservesGroup\PYZsq{}\PY{o+ow}{)} \PY{o+ow}{=}
        \PY{k+kt}{MkHom} \PY{o+ow}{@\PYZob{}}adict\PY{o+ow}{\PYZcb{}} \PY{o+ow}{@\PYZob{}}cdict\PY{o+ow}{\PYZcb{}} \PY{o+ow}{(\PYZbs{}}x \PY{o+ow}{=\PYZgt{}} h\PYZsq{} \PY{o+ow}{(}h x\PY{o+ow}{))} 
             \PY{o+ow}{(\PYZbs{}}something, another \PY{o+ow}{=\PYZgt{}}
                 \PY{o+ow}{(}h\PYZsq{} \PY{o+ow}{(}h \PY{o+ow}{(}something \PY{o+ow}{\PYZlt{}+\PYZgt{}} another\PY{o+ow}{)))}   \PY{o+ow}{=}\PY{o+ow}{\PYZob{}} \PY{o+ow}{?}prf1 \PY{o+ow}{\PYZcb{}=}
                 \PY{o+ow}{(}h\PYZsq{} \PY{o+ow}{(}h something  \PY{o+ow}{\PYZlt{}+\PYZgt{}} h another\PY{o+ow}{))}  \PY{o+ow}{=}\PY{o+ow}{\PYZob{}} \PY{o+ow}{?}prf2 \PY{o+ow}{\PYZcb{}=}
                 \PY{o+ow}{(}h\PYZsq{} \PY{o+ow}{(}h something\PY{o+ow}{)} \PY{o+ow}{\PYZlt{}+\PYZgt{}} h\PYZsq{} \PY{o+ow}{(}h another\PY{o+ow}{))} \PY{k+kt}{QED}\PY{o+ow}{)}
\end{Verbatim}
\caption{Idris equational reasoning.}\label{fig:Equational}
\end{figure}

\begin{figure}[h]
\begin{Verbatim}[commandchars=\\\{\}]
\PY{n+nf}{\PYZgt{}} \PY{o+ow}{:}search \PY{k+kt}{Nat} \PY{o+ow}{\PYZhy{}\PYZgt{}} \PY{k+kt}{Bool}
\PY{o+ow}{=} \PY{k+kt}{Prelude}\PY{o+ow}{.}\PY{k+kt}{Nat}\PY{o+ow}{.}isSucc \PY{o+ow}{:} \PY{k+kt}{Nat} \PY{o+ow}{\PYZhy{}\PYZgt{}} \PY{k+kt}{Bool}
\PY{o+ow}{=} \PY{k+kt}{Prelude}\PY{o+ow}{.}\PY{k+kt}{Nat}\PY{o+ow}{.}isZero \PY{o+ow}{:} \PY{k+kt}{Nat} \PY{o+ow}{\PYZhy{}\PYZgt{}} \PY{k+kt}{Bool}
\end{Verbatim}
\caption{Idris proof search.}\label{fig:ReplSearch}
\end{figure}

These two features, together with broader type libraries, utility lemmas and improved error reporting should go a long to way to making PDD a viable secure software construction method.

\subsection{Design By Contract}

Meyer, in his pioneering work on Design By Contract (DBC) \cite{Meyer:DBC:92}, was one of the first to realize this logical extension of Hoare logic (preconditions, invariants and postconditions).  Proof driven development extends this idea further.  
The limitations of the language's type system define the limits of the assertions that can be made within the contracts.
These limitations include being limited to existential assertions, and the capabilities of statically typed systems with weaker foundations than dependent types.   If previous systems like Java, C\#, C++, C, Python, Haskell, etc. supported dependent types, design by contact frameworks would be capable of encoding very nearly the same types of assertions (both existentially and universally quantified).

This class of limitations applies regardless of the ``level" at which the predicate is tested.  Specifically, in DBC, three levels are available: preconditions, postconditions and (loop-)invariants.  At each level the same limitations of what properties can be asserted exist.  For example: Java types cannot define functional types, so a method cannot take another method as an argument; also, Haskell's type system (including algebraic data types) can only parameterize based on existing top-level types (the full details of this are available in Weirich's work on dependent type systems \cite{Weirich:DependentTypes:12}). 



\section{Examples}

In this section, we show examples of the issues that arise following a ``common" approach.  We illustrate the changes to design and how we can utilize the compiler and error messages in a ``PDD" approach.

The challenges that arise from a standard approach are discussed in section \ref{nonPDDFlow}.  The alternative approach based on PDD is discussed in section \ref{pddFlow}.  In order to provide a valid comparison, we will show issues with creating a function to parse characters into integers.
For example, the character '1' represents the integer 1, and should successfully parse.  On the other hand, the character 'N' does not represent a valid integer and should not parse.

\subsection{Common Approach}

In this section we will work through the creation of a function and associated proof in the ``common" workflow.  

In Idris, the notion of parsing a character that may or may not be valid can be captured with the definition in figure \ref{fig:ParseIntBody}.  We look for the 10 valid cases, and return Nothing if it does not match any of them.  (Clearly this code can be improved, but for illustration purposes it is clear that it is correct).

\begin{figure}[h]
\begin{Verbatim}[commandchars=\\\{\}]
\PY{n+nf}{parseInteger} \PY{o+ow}{:} \PY{k+kt}{Char} \PY{o+ow}{\PYZhy{}\PYZgt{}} \PY{k+kt}{Maybe} \PY{k+kt}{Int}
parseInteger \PY{l+s+sc}{\PYZsq{}}\PY{l+s+sc}{0}\PY{l+s+sc}{\PYZsq{}} \PY{o+ow}{=} \PY{k+kt}{Just} \PY{l+m+mi}{0} 
parseInteger \PY{l+s+sc}{\PYZsq{}}\PY{l+s+sc}{1}\PY{l+s+sc}{\PYZsq{}} \PY{o+ow}{=} \PY{k+kt}{Just} \PY{l+m+mi}{1}
parseInteger \PY{l+s+sc}{\PYZsq{}}\PY{l+s+sc}{2}\PY{l+s+sc}{\PYZsq{}} \PY{o+ow}{=} \PY{k+kt}{Just} \PY{l+m+mi}{2} 
parseInteger \PY{l+s+sc}{\PYZsq{}}\PY{l+s+sc}{3}\PY{l+s+sc}{\PYZsq{}} \PY{o+ow}{=} \PY{k+kt}{Just} \PY{l+m+mi}{3} 
parseInteger \PY{l+s+sc}{\PYZsq{}}\PY{l+s+sc}{4}\PY{l+s+sc}{\PYZsq{}} \PY{o+ow}{=} \PY{k+kt}{Just} \PY{l+m+mi}{4} 
parseInteger \PY{l+s+sc}{\PYZsq{}}\PY{l+s+sc}{5}\PY{l+s+sc}{\PYZsq{}} \PY{o+ow}{=} \PY{k+kt}{Just} \PY{l+m+mi}{5} 
parseInteger \PY{l+s+sc}{\PYZsq{}}\PY{l+s+sc}{6}\PY{l+s+sc}{\PYZsq{}} \PY{o+ow}{=} \PY{k+kt}{Just} \PY{l+m+mi}{6} 
parseInteger \PY{l+s+sc}{\PYZsq{}}\PY{l+s+sc}{7}\PY{l+s+sc}{\PYZsq{}} \PY{o+ow}{=} \PY{k+kt}{Just} \PY{l+m+mi}{7} 
parseInteger \PY{l+s+sc}{\PYZsq{}}\PY{l+s+sc}{8}\PY{l+s+sc}{\PYZsq{}} \PY{o+ow}{=} \PY{k+kt}{Just} \PY{l+m+mi}{8} 
parseInteger \PY{l+s+sc}{\PYZsq{}}\PY{l+s+sc}{9}\PY{l+s+sc}{\PYZsq{}} \PY{o+ow}{=} \PY{k+kt}{Just} \PY{l+m+mi}{9}
parseInteger \PY{k+kr}{\PYZus{}}   \PY{o+ow}{=} \PY{k+kt}{Nothing}
\end{Verbatim}
\caption{Idris type declaration for parsing integers.}\label{fig:ParseIntBody}
\end{figure}

We would like to prove, then, that any character not between '0' and '9' yields Nothing.  Suppose we have a function ``isNumeric" which takes a Char and returns a Boolean (indicating whether the character is between '0' and '9').  Then, we can define the statement of such a proof as shown in figure \ref{fig:ParseIntProofStmt}.

\begin{figure}[h]
\begin{Verbatim}[commandchars=\\\{\}]
\PY{n+nf}{invalidCharAlwaysNothing} \PY{o+ow}{:} \PY{o+ow}{(}c \PY{o+ow}{:} \PY{k+kt}{Char}\PY{o+ow}{)} \PY{o+ow}{\PYZhy{}\PYZgt{}} \PY{o+ow}{(}pf \PY{o+ow}{:} isNumeric c \PY{o+ow}{=} \PY{k+kt}{False}\PY{o+ow}{)} 
                           \PY{o+ow}{\PYZhy{}\PYZgt{}} parseInteger c \PY{o+ow}{=} \PY{k+kt}{Nothing}
invalidCharAlwaysNothing c pf \PY{o+ow}{=} \PY{o+ow}{?}invalidCharAlwaysNothing\PYZus{}rhs
\end{Verbatim}
\caption{Idris proof statement.}\label{fig:ParseIntProofStmt}
\end{figure}

\begin{figure}[h]
\begin{Verbatim}[commandchars=\\\{\}]
\PY{n+nf}{parse\PYZgt{}} \PY{o+ow}{:}p invalidCharAlwaysNothing\PYZus{}rhs 
\PY{c+c1}{\PYZhy{}\PYZhy{}\PYZhy{}\PYZhy{}\PYZhy{}\PYZhy{}\PYZhy{}\PYZhy{}\PYZhy{}\PYZhy{}              Assumptions:              \PYZhy{}\PYZhy{}\PYZhy{}\PYZhy{}\PYZhy{}\PYZhy{}\PYZhy{}\PYZhy{}\PYZhy{}\PYZhy{}}
\PY{n+nf}{c} \PY{o+ow}{:} \PY{k+kt}{Char}
\PY{n+nf}{pf} \PY{o+ow}{:} isNumeric c \PY{o+ow}{=} \PY{k+kt}{False}
\PY{c+c1}{\PYZhy{}\PYZhy{}\PYZhy{}\PYZhy{}\PYZhy{}\PYZhy{}\PYZhy{}\PYZhy{}\PYZhy{}\PYZhy{}                 Goal:                  \PYZhy{}\PYZhy{}\PYZhy{}\PYZhy{}\PYZhy{}\PYZhy{}\PYZhy{}\PYZhy{}\PYZhy{}\PYZhy{}}
\PY{o+ow}{\PYZob{}}hole3\PY{o+ow}{\PYZcb{}} \PY{o+ow}{:} parseInteger c \PY{o+ow}{=} \PY{k+kt}{Nothing}
\end{Verbatim}
\caption{Idris proof status.}\label{fig:ParseIntProofStatus}
\end{figure}

When we attempt to prove the statement, the problem becomes intractable (figure \ref{fig:ParseIntProofStatus}). This is because there is nothing to connect the implementation of ``parseInteger" to the stipulated portion of the proof, specifically that the character is numeric.   To move forward, we must examine the definition of our parse function, and change it to use our numeric query function, or similarly sever the conditional logic.  This happens after 10 lines of very obvious implementation code.  The problem grows with larger code segments.  

As an example of a larger code segment, during string concatention, if we allow for ``negative characters" (such that 'a' concatenated with its negative would yield the empty string) proving even a simple property such as that a string concatenated with the empty string becomes extremely involved.  The proof state corresponding to this is shown in figure \ref{fig:ComplexProofState}.  (The full details of such string manipulation are available online\footnote{https://github.com/bgoodspeed/idris-strings}).

\begin{figure}[h]
\begin{Verbatim}[commandchars=\\\{\}]
\PY{c+c1}{\PYZhy{}\PYZhy{}\PYZhy{}\PYZhy{}\PYZhy{}\PYZhy{}\PYZhy{}\PYZhy{}\PYZhy{}\PYZhy{}              Assumptions:              \PYZhy{}\PYZhy{}\PYZhy{}\PYZhy{}\PYZhy{}\PYZhy{}\PYZhy{}\PYZhy{}\PYZhy{}\PYZhy{}}
 \PY{n+nf}{c} \PY{o+ow}{:} \PY{k+kt}{SignedChar}
 \PY{n+nf}{w} \PY{o+ow}{:} \PY{k+kt}{Word}
 \PY{n+nf}{inductiveHypothesis} \PY{o+ow}{:} wordConcatAndCollapse w \PY{k+kt}{Empty} \PY{o+ow}{=} w
\PY{c+c1}{\PYZhy{}\PYZhy{}\PYZhy{}\PYZhy{}\PYZhy{}\PYZhy{}\PYZhy{}\PYZhy{}\PYZhy{}\PYZhy{}                 Goal:                  \PYZhy{}\PYZhy{}\PYZhy{}\PYZhy{}\PYZhy{}\PYZhy{}\PYZhy{}\PYZhy{}\PYZhy{}\PYZhy{}}
\PY{o+ow}{\PYZob{}}hole3\PY{o+ow}{\PYZcb{}} \PY{o+ow}{:} \PY{k+kr}{case} block \PY{k+kr}{in} 
   wordCollapse \PY{o+ow}{(}c \PY{o+ow}{\PYZsh{}} wordConcat w \PY{k+kt}{Empty}\PY{o+ow}{)}
                \PY{o+ow}{(}wordCollapseOneLevel \PY{o+ow}{(}c \PY{o+ow}{\PYZsh{}} wordConcat w \PY{k+kt}{Empty}\PY{o+ow}{))}
   \PY{o+ow}{(}\PY{k+kt}{Words}\PY{o+ow}{.}\PY{k+kt}{Word} \PY{k+kr}{instance} \PY{k+kr}{of} \PY{k+kt}{Prelude}\PY{o+ow}{.}\PY{k+kt}{Classes}\PY{o+ow}{.}\PY{k+kt}{Eq}, 
       method \PY{o+ow}{=}\PY{o+ow}{=} \PY{o+ow}{(}c \PY{o+ow}{\PYZsh{}} wordConcat w \PY{k+kt}{Empty}\PY{o+ow}{)}
                 \PY{o+ow}{(}wordCollapseOneLevel 
                    \PY{o+ow}{(}c \PY{o+ow}{\PYZsh{}} wordConcat w \PY{k+kt}{Empty}\PY{o+ow}{)))} \PY{o+ow}{=} c \PY{o+ow}{\PYZsh{}} w
\end{Verbatim}
\caption{Idris complex proof status.}\label{fig:ComplexProofState}
\end{figure}

\label{nonPDDFlow}

\subsection{PDD Approach}

As shown in figure \ref{fig:ParseIntProofStmt}, if we approach the same problem with a proof statement first but before creating the code, the necessary connection between the qualifier (isNumeric) and the implementation is more obvious.  In fact, we can now encode the relationship directly into the datatype (figure \ref{fig:DatatypeDecl}).

\begin{figure}[h]
\begin{Verbatim}[commandchars=\\\{\}]
\PY{k+kr}{data} \PY{k+kt}{NumChar} \PY{o+ow}{:} \PY{k+kt}{Type}  \PY{k+kr}{where}
  \PY{n+nf}{MkNumChar} \PY{o+ow}{:} \PY{o+ow}{(}c \PY{o+ow}{:} \PY{k+kt}{Char}\PY{o+ow}{)} \PY{o+ow}{\PYZhy{}\PYZgt{}} \PY{o+ow}{(}ok\PY{o+ow}{:} isNumeric c \PY{o+ow}{=} \PY{k+kt}{True}\PY{o+ow}{)} \PY{o+ow}{\PYZhy{}\PYZgt{}} \PY{k+kt}{NumChar}
   
\end{Verbatim}
\caption{Idris data type declaration.}\label{fig:DatatypeDecl}
\end{figure}

This means we can declare the signature of our ``parseInteger" function using the new datatype (figure \ref{fig:ParseIntDTA}).  Using this, we realize that we cannot even construct a call to ``parseInteger" using a non-numeric character.  The compiler refuses to unify ``False = True" (figure \ref{fig:CompilerError}).  This means we can remove the possibility of failure from the method signature, removing the Maybe clause (figure \ref{fig:ParseIntDTB}).

\begin{figure}[h]
\begin{Verbatim}[commandchars=\\\{\}]
\PY{n+nf}{parseInteger} \PY{o+ow}{:} \PY{k+kt}{NumChar} \PY{o+ow}{\PYZhy{}\PYZgt{}} \PY{k+kt}{Maybe} \PY{k+kt}{Int}
\end{Verbatim}
\caption{Idris improved parseInteger declaration (A).}\label{fig:ParseIntDTA}
\end{figure}

\begin{figure}[h]
\begin{Verbatim}[commandchars=\\\{\}]
parse\PY{o+ow}{\PYZgt{}} parseInteger \PY{o+ow}{(}\PY{k+kt}{MkNumChar} \PY{l+s+sc}{\PYZsq{}}\PY{l+s+sc}{c}\PY{l+s+sc}{\PYZsq{}} \PY{k+kt}{Refl}\PY{o+ow}{)}
\PY{o+ow}{(}input\PY{o+ow}{):}\PY{l+m+mi}{1}\PY{o+ow}{:}\PY{l+m+mi}{25}\PY{o+ow}{:}\PY{k+kt}{When} elaborating argument ok 
        to constructor \PY{k+kt}{Main}\PY{o+ow}{.}\PY{k+kt}{MkNumChar}\PY{o+ow}{:}
        \PY{k+kt}{Can\PYZsq{}t} unify
                x \PY{o+ow}{=} x
        \PY{k+kr}{with}
                isNumeric \PY{l+s+sc}{\PYZsq{}}\PY{l+s+sc}{c}\PY{l+s+sc}{\PYZsq{}} \PY{o+ow}{=} \PY{k+kt}{True}
        
        \PY{n+nf}{Specifically}\PY{o+ow}{:}
                \PY{k+kt}{Can\PYZsq{}t} unify
                        \PY{k+kt}{False}
                \PY{k+kr}{with}
                        \PY{k+kt}{True}
\end{Verbatim}
\caption{Idris compilation error for bad invocation.}\label{fig:CompilerError}
\end{figure}

\begin{figure}[h]
\begin{Verbatim}[commandchars=\\\{\}]
\PY{n+nf}{parseInteger} \PY{o+ow}{:} \PY{k+kt}{NumChar} \PY{o+ow}{\PYZhy{}\PYZgt{}} \PY{k+kt}{Int}
\end{Verbatim}
\caption{Idris improved parseInteger declaration (B).}\label{fig:ParseIntDTB}
\end{figure}

\label{pddFlow}

This small change in workflow for a trivial function has many improvements to the resulting design.  This is true not only to the design of the code developed here, but also for any client code making use of the function.  Rather than using a parse routine and checking afterwards to see if something went wrong, it is now up to the client to prove nothing will go wrong \textit{before it is allowed to make the call}.


\section{Impacts}

As discussed in the previous section, the use of PDD extends TDD (there are assertions that can be made in a proof-driven environment that cannot be made in a test-driven environment) and permits universal quantification of system properties.  The use of proof assistants in general close an important security gap, as shown in figures \ref{fig:OriginalNumberedWorkflow}
  and \ref{fig:ImprovedNumberedWorkflow}.

Just as TDD was shown to alter the structure of the code under test/development (by making hooks for tests to assert intermediate results), so too does PDD.  In this case, we have intermediate placeholders for assertions, as well as intermediate dependent data types (which carry their properties with them, similar to invariants in Hoare logic \cite{Hoare:AxiomaticBasis:69}).

It is possible for a proof script itself to contain errors, thus ``locking in" erroneous behavior in the system.  Of course, this is true of ink and paper proofs as well.  The traditional remedy for such paper proofs is peer review.  The corresponding software quality technique is code review, which has been shown to positively impact quality \cite{Kemerer:CodeReview:09}.  This is sometimes done during development, as in pair programming used by the XP \cite{Beck:XPExplained:99} process, and sometimes after the system is completed.  The benefits arising from code reviews happening after construction is that the system is viewed by someone not involved in its creation.  This has had a substantial impact on the quality of systems as a whole \cite{Kemerer:CodeReview:09, Mantyla:DefectsFromReviews:09}.

It should be noted that this is not a `silver bullet' \cite{Brooks:Mythical:75}.  It should be easier to prove properties about code created with proofs first than trying to prove properties post-hoc.  The extreme cost and difficulties associated with machine verified proofs are is also worth noting.  In one successful system, the seL4 verified microkernel (a formally verified operating system kernel) \cite{Klean:FormalKernel:09}, development was completed in approximately 2 person-months, but verification efforts took 20 person-years \cite{Ricketts:ProofAssistant:14}.   Substantial work is required to make verified software construction easier for industrial users as well as expert academics before it can become mainstream.

\section{Related Work}
We  saw the formalization of the processes used to create software in the late 1990s and early 2000s \cite{Beznosov:AgileSecAssurance:04, Beck:TDD:03,  Schwaber:SCRUM:01, Beck:XPExplained:99}. 

Quality assurance, for our purposes, is the act of verifying that an implemented system (per the workflow described in section \ref{workflow}) meets the goals or policies defined by the specification phase.

This is often done by testing.  Fundamentally, there is a divide between manual  testing (where a person operates the machine/runs the program) and automated testing (where the machine consults an oracle \cite{Miller:OracleTesting:78}  to determine the correct results).  Formally, procedures (such as ISO 9000 \cite{Tsim:ISO9000:02}, clean room \cite{Prowell:CleanRoom:99}, and waterfall \cite{Benington:Waterfall:83}) have been used to facilitate manual testing and apply rigor to a human endeavor.  Semi-recently, agile software development \cite{Martin:CleanCode:08, Beznosov:AgileSecAssurance:04,Schwaber:SCRUM:01, Beck:XPExplained:99}  has pushed for new processes like test-driven development (TDD) \cite{Beck:TDD:03} to shift more of the work onto the computer.  This has had a positive effect on defect rates, in some cases reducing the number of defects by 50\% \cite{Maximilien:TDDDefectRates:03}.

The theoretical framework of a programming language defines the limits of what can be described by the primitives of the language.  In the earliest languages (assembly languages), the framework was dictated by the logic gate structures of the hardware on which they ran  \cite{Knuth:EarlyLanguages:09}.  Early higher level languages (ones that require translation or compilation to run as machine code), such as LISP and Fortran, were based on very early formalisms of the lambda calculus\cite{Church:LambdaCalc:41} and Turing machines \cite{Turing:ComputableNumbers:37}.  More recent advances in programming language theory and type theory gave rise to systems like the calculus of inductive constructions \cite{Coquand:CalcConstruct:86}, 
 which allowed us to build systems like Coq \cite{Bertot:CoqArt:04} and Isabelle \cite{ Paulson:Isabelle:89}.  Martin-Lof type theory \cite{MartinLof:TypeTheory:84} gave rise to dependent types \cite{Chlipala:DependentTypes:08, McKinna:WhyDependent:06, Xi:PracticalDepTypes:99}, which form the basis of the systems  Agda \cite{Norell:Agda:09}, Idris \cite{Brady:IdrisSystemsProgramming:11} and Cayenne \cite{Augustsson:Cayenne:99}.

\section{Conclusion}

We demonstrated that our new workflow, PDD, can express universal quantification.  This is broader in scope than existential quantification as used in TDD, which itself had desirable properties in the SDLC.    We analysed the available dependently typed systems, and justified our choice of Idris for this research.  We have used the lessons of other recent developments with related languages and for related goals to avoid re-inventing the wheel.

While several steps have been taken, some important issues of difficulty and utility of these systems remain to be solved before widespread industrial use is likely.  We discuss several avenues of future work below in the hopes that some of these can be addressed in the future.

\section{Future Work}

Dependently typed programming is a young discipline, and the paradigm is not yet well understood.  The implications of design decisions are not clear, nor are best practices.  Much experimentation will be required before we can tell the long term ramifications of our type designs.

\subsection{Error Messages}

As with TDD, the quality of error messages determines the utility of the workflow - if we want the process to guide us towards correct code, we need to know where we've gone off the rails.  With the availability of proof search, and the compiler ability to query for the type required to fill a hole in a program, it should be possible to generate suggestions within the system library as to: a) the exact signature required of a method to complete the program; and, b) known methods that have the required signature.  A more flexible search that can consider composite expressions or interim datatypes would be useful.

When proofs fail to unify (as when assertions fail to pass), it would be extremely helpful to have a return value saying why and exactly where it failed.  

In particular, Idris does not inform the user during interactive proofs that a rewrite has failed.  The compiler silently fails to change the target expression. 
\subsection{Proof Targeting}
 It would be useful to support pre-order reasoning (as in equational reasoning in mathematical proof), so that intermediate transformations can happen.  This would more closely mimic the way mathematics are done without computer assistance, and would therefore be a more familiar and gentler introduction to mathematical verification.

\section{Acknowledgements} 
Thank you to Dr. Konstantinidis for his help with this research.

\bibliographystyle{unsrt}  
\bibliography{research}  

\end{document}